\documentclass{aa}
\usepackage{graphicx}
\begin{document}

\title{The interferometric diameter and internal structure of Sirius~A}

\author{P. Kervella\inst{1},
F.~Th\'evenin\inst{2},
P.~Morel\inst{2},
P.~Bord\'e\inst{3}
\and
E.~Di Folco\inst{4}}
\offprints{P. Kervella}

\institute{European Southern Observatory, Alonso de Cordova 3107, Casilla 19001, Vitacura, Santiago 19, Chile
\and D\'epartement Cassini, UMR CNRS 6529, Observatoire de la C\^ote 
   d'Azur, BP 4229, 06304 Nice Cedex 4, France
\and LESIA, Observatoire de Paris-Meudon, 5, place Jules Janssen, F-92195 Meudon Cedex, France
\and European Southern Observatory, Karl-Schwarzschild-str. 2, D-85748 Garching, Germany}

\titlerunning{The diameter and structure of Sirius~A}
\authorrunning{P. Kervella et al.}
\mail{pkervell@eso.org}

\date{Received ; Accepted }

\abstract{
The interferometric observations of dwarf stars in the solar neighbourhood,
combined with {\sc Hipparcos} parallaxes provide very precise values
of their linear diameters. In this paper, we report the direct measurement of the
angular diameter of the bright star Sirius~A with the VINCI/VLTI instrument.
We obtain a uniform disk angular diameter of
$\theta_{\rm UD} = 5.936\, \pm \, 0.016$~mas in the $K$ band
and a limb darkened value of
$\theta_{\rm LD} = 6.039 \, \pm \, 0.019$~mas.
In combination with the {\sc Hipparcos} parallax of 379.22 $\pm$ 1.58~mas,
this translates into a linear diameter of 1.711~$\pm$~0.013~D$_{\odot}$.
Using the VINCI/VLTI interferometric diameter and the published properties of
Sirius~A, we derive internal structure models corresponding to
ages between $200$ and $250 \pm 12$~Myr. This range is defined mainly
by the hypothesis on the mass of the star, the overshoot and the metallicity. 
}
\maketitle
\section{Introduction}
The brightest star in the sky, Sirius~A (\object{HR 2491}, \object{HD 48915}),
is an A1V type dwarf. It is also a member of the fifth nearest binary system.
The existence of a massive companion, Sirius~B, was predicted by Bessel~(\cite{bessel1844})
based on the apparent motion of Sirius~A. It was later discovered visually by A. G. Clark
in 1862, and is still today among the most massive white dwarfs (WD) known.
Chandrasekhar (\cite{chandra35}) published the first model to explain its
characteristics. Sirius has been reported several times to be a red star
in the ancient times, whereas it is now almost white. The reason behind this
color change is still not understood, though several scenarios have been proposed
in the last decades (Schlosser \& Bergmann~\cite{schlosser85},
Bonnet-Bideau \& Gry~\cite{bonnet91}, Whittet~\cite{whittet99}).
During the past century, many efforts were directed towards compiling the 
properties of both stars like their masses or chemical abundances. Sirius~A exhibits
enhanced metal abundances in its  photosphere and has therefore been classified
as Am by Abt~(\cite{abt79}).
As the orbital period is relatively short (50 years), two revolutions of the
Sirius~A-B binary have already been measured, allowing
Gatewood \& Gatewood~(\cite{gatewood78}) to derive
precisely the masses of the two components (Sect.~\ref{masses}).
Moreover, Sirius~A  has been the first main sequence star to have its diameter
measured interferometrically by Hanbury Brown et al.~(\cite{hanbury67},
\cite{hanbury74a}).

The chief object in our study is to model the internal structure of 
Sirius~A in order to reproduce its macroscopic characteristics. In this process,
we use the angular diameter of this star from our new VINCI/VLTI measurements
and the {\sc Hipparcos} parallax to derive its linear diameter.
From our models, we compute the age and the initial helium content of Sirius~A.
We summarize in Sect.~\ref{fund_par} the fundamental parameters of this star, while
Sect. \ref{data_proc} to \ref{angdiam} are dedicated to the VINCI data processing
and analysis. This section includes in particular the fit of the limb darkened disk
model to the measured visibilities.
Finally, we discuss in Sect.~\ref{modeling} the macroscopic constraints
that we apply to our CESAM numerical models (Morel \cite{morel97}) of the internal
structure of Sirius~A, and we predict the asteroseismic large frequency spacing of this star.
\section{Fundamental stellar parameters \label{fund_par}}
\subsection{Masses \label{masses}}
Sirius~(A1V + WD) is a visual binary system with a 50 years period
that has been studied extensively, and its orbit is known with high precision.
From photometric astrometry covering a 60 year period,
Gatewood \& Gatewood~(\cite{gatewood78}) determined the orbital
photocentric semi-major axis of Sirius A with respect to the system barycenter.
This determination relied upon the orbital parameters of the relative
visual orbit of Sirius~B given by van~den~Bos (\cite{vandenbos60}).
These authors derived a mass for Sirius~A of 2.143 $\pm$ 0.056 M$_{\odot}$,
using a parallax of $\pi = 377.7 \pm 3.31$~mas.
This precision of $\pi$ has been recently improved thanks to
the {\sc Hipparcos} satellite  mission (Perryman et al.~\cite{hip})
at $\pi = 379.22 \pm 1.58$~mas. We note that this value is consistent
with the parallax of Van Altena et al.\,(\cite{vanaltena95}),
$\pi = 381.6 \pm 2.2$\,mas.
Consequently, the sum of the masses has to be diminished by 1.2\%,
giving for Sirius~A a lower mass of 2.12 $\pm$ 0.06~M$_{\odot}$.
Independently of binary orbital elements, 
Holberg et al.~(\cite{holberg98}) have re-determined
the mass of Sirius~B from  precise measurements of its surface
gravity and radius. Using the {\sc Hipparcos} parallax they found
a mass of M$_{\rm B} = 1.034 \pm 0.026$\,M$_{\odot}$ in
agreement with the revisited sum of the masses.
Also based on the {\sc Hipparcos} parallax,
Provencal et al.~(\cite{provencal98}) have derived a value of
M$_{\rm B} = 1.000 \pm 0.016$\,M$_{\odot}$, in agreement with Holberg et
al.~(\cite{holberg98}) at a 1.1\,$\sigma$ level. 
\subsection{Age}

The age of the Sirius system has been poorly debated in the literature.
Based on Wood's (\cite{wood95}) sequences of cooling age of WD
(carbon-oxygen model with thick H layer),
Sirius B is found to be a young WD of about 160 (Holberg et
al.\,\cite{holberg98}) to $210 \pm 20$\,Myr (Holberg, {\it private communication}).
The revisited mass of Sirius~B (1.034~M$_{\odot}$)  corresponds
to a progenitor mass of $\approx 7$\,M$_{\odot}$, i.e.
the approximate mass of a B5 main sequence star,
from the empirical initial to final mass relation
of Weidemann~(\cite{weidemann02}).
Such a massive star evolves to a WD in $40 \pm 5$\,Myr.
Consequently, assuming a simultaneous origin of both A \& B components,
the Sirius system is expected to be between 200 and 250\,Myr old.
\subsection{Spectral properties}
Sirius shows deep metallic spectral lines, and we propose that the corresponding enhancement
in metal abundance is a consequence of the strong radiative forces at work in the interior of this star.
This hypothesis is supported by the very low rotational velocity (Royer et al. \cite{royer02})
$v\sin i \approx 16$ km s$^{-1}$, that favors the action of these forces.
In their comparison with the star Vega, Qiu et al.~(\cite{qiu01}) concluded that
Sirius~A appears to be enhanced in abundances by about 1.0 dex on
average while C, O, Sc and Ca are underabundant.
In consequence, for such a slow rotator and intermediate mass,
it is very hazardous to estimate the mean metal content in the interior of the
star from its surface metal abundance.

Therefore, we propose to focus only on the abundances of CNO,
that are believed to be little modified by radiative forces (Richer et al.~\cite{richer00}),
to estimate the value of $\left(\frac ZX\right)_{\rm s}$ at present day.
We adopt an average abundance at present day of
$\left(\frac ZX\right)_{\rm s}~=~0.012~\pm~0.002$ in our models, that includes the
microscopic diffusion of the elements.
$\left(\frac ZX\right)_{\rm i}$ is very important for the internal stellar opacity of the
computed model and is a key parameter with respect to the age
of the star (see Section \ref{modeling}). 
It results, when using these elements as tracers, that the average metal
abundance of Sirius~A is in fact about half that of the Sun.
While appearing to be low, the adopted $Z_{\rm s}$
abundance is comparable with Vega's observed value. Vega is a fast
rotator and therefore should not be affected by radiative accelerations
masking its true metal abundance.
This corresponds to an initial value at zero age of $\left(\frac ZX\right)_{\rm i}~=~0.0165~\pm~0.002$.
In order to validate the previous assumption that the metallicity of outerlayers
does not represent the metallicity of the whole star, we also present, in Sect.~\ref{modeling},
a model of Sirius A with the observed metallicity $\left(\frac ZX\right)_{\rm s} = 0.0329$,
corresponding to the observed mean surface abundances.
\subsection{Photometry}
The photometric properties of Sirius are summarized in Table~\ref{params}. 
The magnitude and the parallax give an intrinsic luminosity of
$L/L_{\odot}=25.4\pm1.3$. Following Lemke~(\cite{lemke89}),
we adopt an effective temperature of $T_{\rm eff}$ = 9900 $\pm$ 200~K and 
a logarithm of the spectroscopic surface gravity of $\log g$ = 4.3 $\pm$ 0.1.

\section{Interferometric observations and data processing \label{data_proc}}
\subsection{Instrumental setup}
The European Southern Observatory's Very Large Telescope Interferometer
(Glindemann et al. \cite{glindemann}) is operated  on top of the Cerro Paranal, in Northern Chile since
March 2001. For the observations reported in this paper, the light coming from two test siderostats
(0.35~m aperture) was recombined coherently in VINCI, the VLT INterferometer Commissioning Instrument
(Kervella et al. \cite{kervella00}).  We used a regular K band filter ($\lambda = 2.0-2.4 \mu$m)
but VINCI can also be operated in the H band ($\lambda = 1.4-1.8 \mu$m) using an
integrated optics beam combiner (Berger et al. \cite{berger01}, Kervella et al. \cite{kervella03a}).
Two VLTI baselines were used for this program: D1-B3 and E0-G1, repsctively 24 and 66m
in ground length.
\subsection{Data processing}
We used a modified version of the standard VINCI data
reduction pipeline (Kervella et al. \cite{kervella03c}), whose general principle
is based on the original FLUOR algorithm (Coud\'e du Foresto et al. \cite{cdf97}).
The two calibrated output interferograms are subtracted to remove residual
photometric fluctuations. Instead of the classical Fourier analysis, we
implemented a time-frequency analysis (S\'egransan et al. \cite{s99})
based on the continuous wavelet transform (Farge \cite{farge92}).
In this approach, the projection of the signal is not done onto a sine wave (Fourier transform),
but on a function, i.e. the wavelet, that is localized in both time and frequency.
We used as a basis the Morlet wavelet, a gaussian envelope
multiplied by a sine wave. With the proper choice of the number of oscillations
inside the gaussian envelope, this wavelet closely matches a VINCI
interferogram. It is therefore very efficient at localizing the signal in both time and frequency.

The differential piston corrupts the amplitude and the shape of the
fringe peak in the wavelets power spectrum. 
A selection based on the shape of fringe peak in the 
time-frequency domain is used to remove ``pistonned'' and false detection interferograms.  
Squared coherence factors $\mu^2$ are then derived by integrating the
wavelet power spectral density (PSD)
of the interferograms at the position and frequency of the fringes.
The residual photon and detector noise are removed
by making a least squares fit of the PSD at high and low frequency.
The resulting measurement stability is satisfactory: on a good series of 500 interferograms of Sirius
(10 minutes) the standard deviation is typically 2.0\%, and the final bootstrapped
statistical error on the average $\mu$ is only 0.16\%.
The $\mu^2$ values are converted into calibrated visibilities $V^2$ through
the observation of calibrator stars (Sect.~\ref{vis_values}).
\begin{table*}

\caption[]{Relevant parameters of $\alpha$\,CMa and its calibrators from the literature.\label{params}}
\begin{tabular}{lcccccc}

Name & $\alpha$\,CMa & $\theta$\,Cen & $\delta$\,Lep & $\alpha$\,Crt & 31\,Ori & HR 4050\\
HD number & \object{HD 48915} & \object{HD 123139} & \object{HD 39364} & \object{HD 95272} & 
\object{HD 36167} & \object{HD 89388}\\
\hline

\noalign{\smallskip}

$m_\mathrm{V}$ & -1.47 & 2.06 & 3.80 & 4.07 & 4.71 & 3.38 \\
$m_\mathrm{K}$ & -1.31 & -0.10 & 1.31 & 1.62 & 0.90 & 0.60\\
Sp. Type & A1V & K0IIIb & G8III/IV & K1III & K5III & K3IIa \\
$\rm T_{\mathrm{eff}}$ (K)$^{\mathrm{a}}$ & 9900  & 4980 & 4580 & 4650 & 3930 & 4500 \\
$[\mathrm{Fe}/\mathrm{H}]^{\mathrm{f}}$ & 0.5 & 0.0 & -0.75 & -0.2 & -0.3 & 0.5 \\
$\log g ^{\mathrm{a}}$ & 4.3 & 2.75 & 2.95 & 2.8 & 1.6 & 1.6 \\
$\pi$ (mas)$^{\mathrm{b}}$ & $379.22 \pm 1.58$ & 53.5 $\pm$ 0.8 & 29.1 $\pm$ 0.6 & 18.7 $\pm$ 1.0 & 7.2 $\pm$ 0.8 & 4.4 $\pm$ 0.5 \\
${\theta_{\rm {LD}}}$ (mas)$^{\mathrm{c}}$ & ${\bf 6.039 \pm 0.019}^{\mathrm{g}}$ & $5.434 \pm 0.020^{\mathrm{e}}$ & 2.63 $\pm$ 0.04 & 2.28 $\pm$ 0.03 & 3.66 $\pm$ 0.06 & 5.23 $\pm$ 0.06 \\
${\theta_{\rm {UD}}}$ (mas)$^{\mathrm{d}}$ & ${\bf 5.936 \pm 0.016}^{\mathrm{g}}$ & $5.305 \pm 0.020^{\mathrm{e}}$ & 2.57 $\pm$ 0.04 & 2.22 $\pm$ 0.03 & 3.55 $\pm$ 0.06 & 5.09 $\pm$ 0.06\\
 \noalign{\smallskip}

\hline

\end{tabular}

\begin{list}{}{}

\item[$^{\mathrm{a}}$] From Cohen et al. (\cite{cohen92}) for $\alpha$\,CMa, and from Cayrel de Strobel et al. (\cite{cayrel}, \cite{cayrel01}) for other stars.
\item[$^{\mathrm{b}}$] Parallaxes from the {\sc Hipparcos} catalogue (Perryman et al. \cite{hip}).
\item[$^{\mathrm{c}}$]  Catalogue values from Cohen et al. (\cite{cohen99}) for other stars.
\item[$^{\mathrm{d}}$] Linear limb darkening coefficients factors from Claret et al.~(\cite{claret95}).
\item[$^{\mathrm{e}}$] $\theta$\,Cen angular diameters from Kervella et al.~(\cite{kervella03b}).
\item[$^{\mathrm{f}}$] Qiu et al. for $\alpha$\,CMa, and Cayrel at al. (\cite{cayrel}) for other stars.
\item[$^{\mathrm{g}}$] Sirius angular diameters from this work (see text for details).
\end{list}

\end{table*}

\section{Calibration}\label{vis_values}
The calibration of Sirius visibilities was achieved using
well-known calibrator stars that have been selected in the Cohen et al. (\cite{cohen99}) catalogue.
The characteristics of the selected calibrators are listed in Table \ref{params}.
The limb-darkened disk (LD) angular diameter of these stars was converted into a uniform disk value
using linear coefficients taken from Claret et al.~(\cite{claret95}). As demonstrated by
Bord\'e et al. (\cite{borde}), the star diameters in the Cohen et al.~(\cite{cohen99})
list have been measured very homogeneously
to a relative precision of approximately 1\% and agree
well with other angular diameter estimation methods.
The interferometric efficiency is computed from the coherence factors obtained on
these calibrators, taking into account the bandwidth smearing effect (Sect.~\ref{smearing})
and a uniform disk angular diameter model. This process yields the calibrated squared visibilities
that are used for the model fit.

In Table~\ref{table_vis}, two error bars are given for each $V^2$ value:
\begin{itemize}
\item one statistical error bar, computed from the dispersion of the visibility values obtained during the observation,
\item one systematic error bar defined by the uncertainty on the knowledge of the calibrator stars angular sizes.
\end{itemize}
While the statistical error can be diminished by repeatedly observing the target, the systematic
error cannot be reduced by averaging measurements obtained using the same calibrator. This
is taken into account in our model fitting by checking {\it a posteriori} that the resulting uncertainty
of the model visibility is larger than the systematic errors of each measured visibility value. This
conservative approach ensures that we are not underestimating the final error bars on the angular
diameter of Sirius.
\begin{table*}
\caption{Sirius squared visibilities.}
\label{table_vis}
\begin{tabular}{lccccl}
\hline
JD & Stations & Baseline (m) & Az.$^{\mathrm{b}}$ & $V^2\ \pm$ Stat. $\pm$ Syst. & Calibrators$^{\mathrm{a}}$\\
\hline

2452682.71653 & D1-B3 & 17.758 & 81.68 & 0.8602 $\pm$ 0.0206 $\pm$ 0.0007 & $\delta$\,Lep, $\alpha$\,Crt\\
2452682.71251 & D1-B3 & 18.139 & 81.45 & 0.8742 $\pm$ 0.0210 $\pm$ 0.0007 & $\delta$\,Lep, $\alpha$\,Crt\\
2452679.71932 & D1-B3 & 18.270 & 81.36 & 0.8769 $\pm$ 0.0313 $\pm$ 0.0023 & $\delta$\,Lep, 31\,Ori, HR\,4050\\
2452682.70813 & D1-B3 & 18.544 & 81.18 & 0.8668 $\pm$ 0.0208 $\pm$ 0.0007 & $\delta$\,Lep, $\alpha$\,Crt\\
2452679.71478 & D1-B3 & 18.684 & 81.08 & 0.8748 $\pm$ 0.0309 $\pm$ 0.0023 & $\delta$\,Lep, 31\,Ori, HR\,4050\\
2452679.71085 & D1-B3 & 19.032 & 80.84 & 0.8627 $\pm$ 0.0300 $\pm$ 0.0022 & $\delta$\,Lep, 31\,Ori, HR\,4050\\
2452679.66856 & D1-B3 & 22.063 & 78.01 & 0.8092 $\pm$ 0.0280 $\pm$ 0.0021 & $\delta$\,Lep, 31\,Ori, HR\,4050\\
2452679.66415 & D1-B3 & 22.301 & 77.68 & 0.8110 $\pm$ 0.0281 $\pm$ 0.0021 & $\delta$\,Lep, 31\,Ori, HR\,4050\\
2452679.65967 & D1-B3 & 22.527 & 77.35 & 0.8048 $\pm$ 0.0278 $\pm$ 0.0021 & $\delta$\,Lep, 31\,Ori, HR\,4050\\
2452361.62629 & E0-G1 & 60.439 & 1.29 & 0.1599 $\pm$ 0.0074 $\pm$ 0.0045 & $\theta$\,Cen\\
2452361.64154 & E0-G1 & 60.574 & 5.16 & 0.1608 $\pm$ 0.0062 $\pm$ 0.0045 & $\theta$\,Cen\\
2452361.57907 & E0-G1 & 61.038 & 169.41 & 0.1472 $\pm$ 0.0085 $\pm$ 0.0041 & $\theta$\,Cen\\
2452358.55861 & E0-G1 & 62.027 & 162.72 & 0.1433 $\pm$ 0.0062 $\pm$ 0.0042 & $\theta$\,Cen\\
2452340.60484 & E0-G1 & 62.148 & 162.07 & 0.1311 $\pm$ 0.0059 $\pm$ 0.0014 & $\theta$\,Cen\\
2452340.57040 & E0-G1 & 63.651 & 155.00 & 0.1243 $\pm$ 0.0063 $\pm$ 0.0014 & $\theta$\,Cen\\
2452340.56629 & E0-G1 & 63.831 & 154.23 & 0.1226 $\pm$ 0.0054 $\pm$ 0.0013 & $\theta$\,Cen\\
2452340.54577 & E0-G1 & 64.678 & 150.65 & 0.1186 $\pm$ 0.0051 $\pm$ 0.0013 & $\theta$\,Cen\\
2452340.54210 & E0-G1 & 64.816 & 150.06 & 0.1152 $\pm$ 0.0052 $\pm$ 0.0013 & $\theta$\,Cen\\
\hline
\end{tabular}
\begin{list}{}{}
\item[$^{\mathrm{a}}$]All calibrators for Sirius were chosen in the Cohen et al. (\cite{cohen99}) catalogue.
\item[$^{\mathrm{b}}$]The azimuth is counted clockwise from North, in degrees (E = 90 deg).
\end{list}
\end{table*}

\begin{figure}[t]
\centering
\includegraphics[bb=0 0 360 288, width=8.5cm]{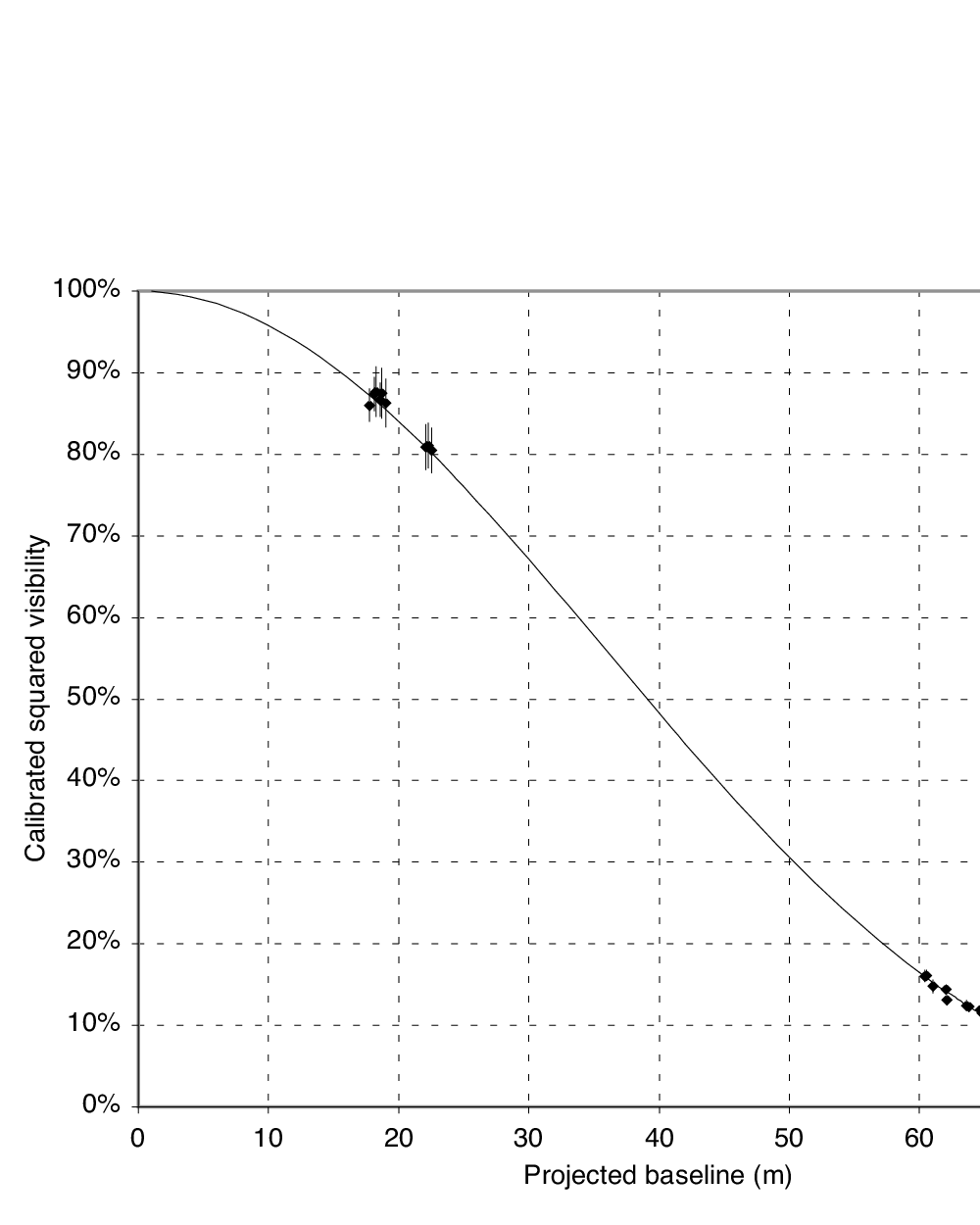}
\caption{Overview of the VINCI/VLTI visibility measurements obtained on Sirius. The solid line 
is the best fit visibility model, that takes into account the limb darkening and the bandwidth smearing.}
\label{sirius_visib_global}
\end{figure}
%
\begin{figure}[t]
\centering
\includegraphics[bb=0 0 360 288, width=8.5cm]{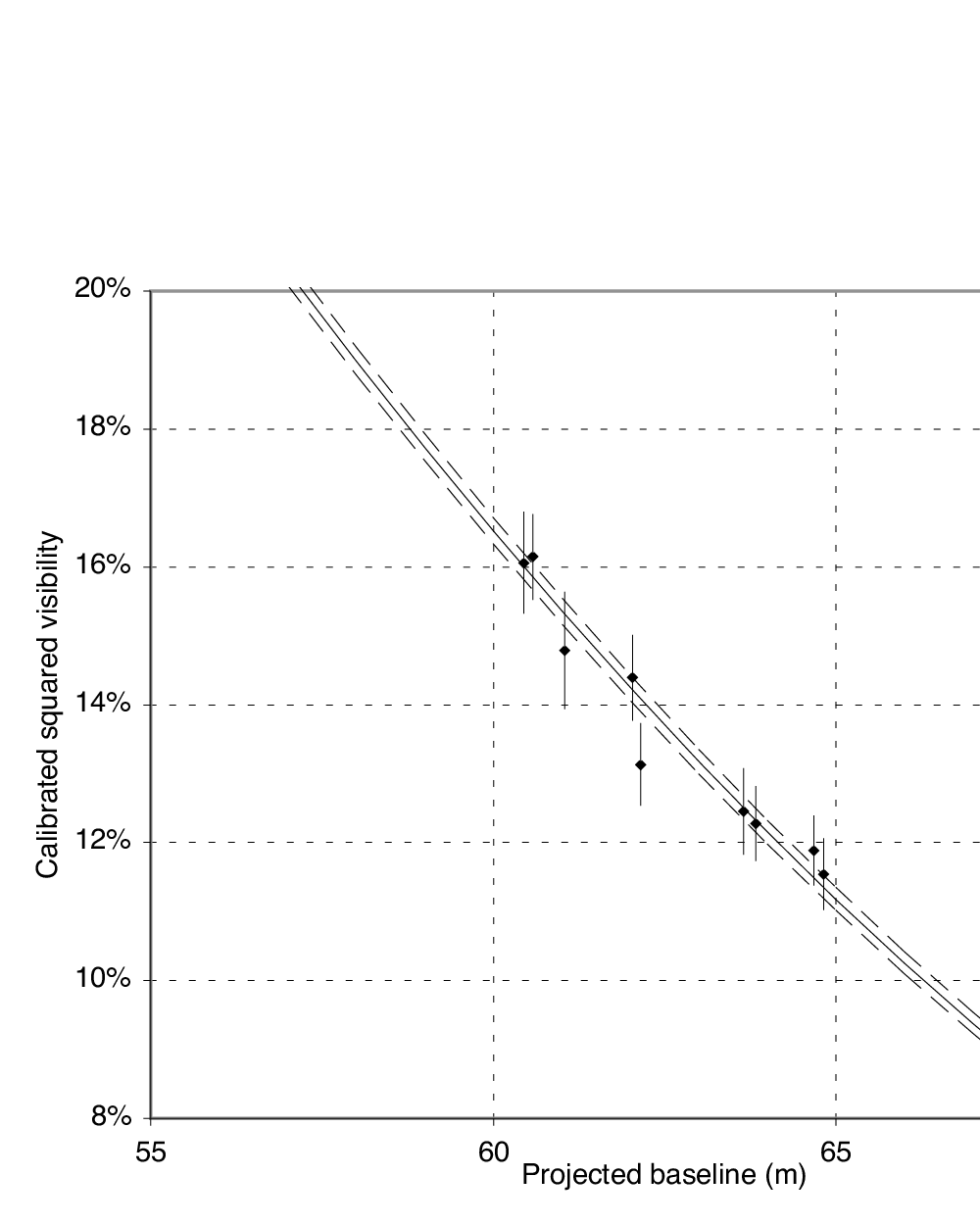}
\caption{Visibility measurements obtained on Sirius on the E0-G1 baseline. The thin line is the best fit model,
and the dashed lines mark the limits of the 1$\sigma$ error domain. For clarity, the plotted error bars are the
quadratic sum of the statistical and systematic errors.}
\label{sirius_visib_detail}
\end{figure}
\section{Angular diameter computation \label{angdiam}}
\subsection{Intensity profile}
From the visibility curve shape before the first minimum, it is almost impossible
to distinguish between a uniform disk (UD) and limb darkened (LD) model. Therefore, it
is necessary to use a model of the stellar disk limb darkening to deduce the photospheric angular
size of the star, from the observed visibility values.
The intensity profiles that we used were computed by Claret~(\cite{claret00}), based on
model atmospheres by Kurucz~(\cite{kurucz92}).
We chose the four parameters $I(\mu)/I(1)$ law of this author, where
$\mu = \cos \theta$ is the cosine of the azimuth of a surface element of the star. They are
accurate approximations of the numerical results from the ATLAS modeling code.

\begin{figure}[t]
\centering
\includegraphics[bb=0 0 360 288, width=8.5cm]{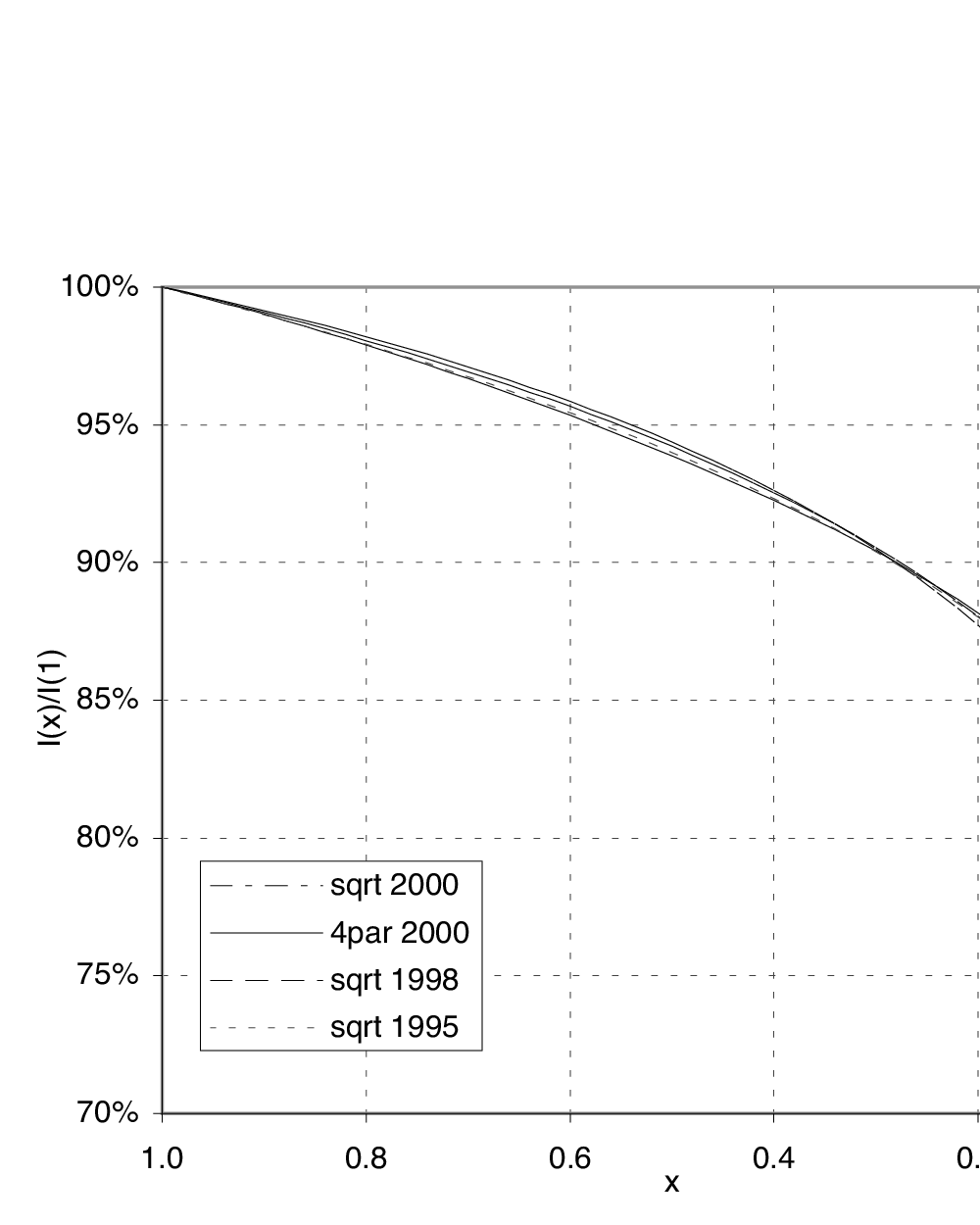}
\caption{Published LD laws for Sirius in the $K$ band, from Claret et al.~(\cite{claret95}),
Claret~(\cite{claret98}) and Claret~(\cite{claret00}).
The square root law is already a good approximation
to the Kurucz atmosphere models. The solid line shows results with
the most recent four parameters law
introduced by Claret~(\cite{claret00}).
This last version was chosen for the Sirius LD fit of this paper.}
\label{limb_darkening_laws}
\end{figure}

To compare the different types of approximated limb darkening laws,
we chose the following approximate parameters for Sirius:
$T_{\rm eff}$~=~10000~K (for Claret et al.~\cite{claret95} and Claret~\cite{claret00})
or $T_{\rm eff}$~=~9800~K (for Claret~\cite{claret98}), $\log g = $~4.0, [Fe/H]~=~+0.5,
$V_{\rm turb} = $~0~km.s$^{-1}$.
These were chosen as close as possible to the Cohen et al.~(\cite{cohen92}) values.
As noted by Claret~(\cite{claret00}), the sensitivity of the limb
darkening of a $T_{\rm eff} = 10000$ K star to the metallicity is negligible,
so the true metal content of Sirius is not critical.
Different $I(\mu)/I(1)$ square root and four parameters laws for Sirius, from Claret et al.~(\cite{claret95})
and Claret (\cite{claret98}, \cite{claret00}) are plotted on Fig.~\ref{limb_darkening_laws}.
Though no formal error bars are provided for these limb darkening models, the small differences between
the four curves demonstrates the good internal consistency of the different types of laws. 

The limb darkening is directly measurable by interferometry beyond the first minimum
of the visibility function, as demonstrated by several authors on giant stars
(Quirrenbach et al.~\cite{quirrenbach96}, Wittkowski et al.~\cite{wittkowski01}).
Unfortunately, for stars of the angular size of Sirius observed in the K band, this requires baseline of 80 to 120
meters that were not available for the measurements reported here. It is intended in the near future
to measure directly the LD of a number of nearby stars, using the VINCI and
AMBER (Petrov et al.~\cite{petrov00}) instruments on the long baselines of the VLTI.
With lengths of up to 202m, they will allow the exploration of the secondary and higher
order lobes of the visibility function.

\subsection{Visibility model \label{smearing}}

The VINCI instrument bandpass corresponds to the $K$ band filter (2-2.4 $\mu$m).
An important effect of this relatively large spectral bandwidth
is that several spatial frequencies are simultaneously
observed by the interferometer. This effect is known as the {\it bandwidth smearing}.
It is usually negligible for $V^2 \ge 40$\%,  but this is not the case when visibilities
come closer to the first minimum of the visibility function.
In the case of Sirius~A observed with the E0-G1
baseline, $V^2$ is already low (~13\%) and we have to consider it.

To account for the bandwidth smearing, the model visibility is computed for regularly spaced
wavenumber spectral bins over the $K$ band, and then integrated to obtain the model visibility.
We assume in this paper a limb darkening model following the four parameters
law of Claret~(\cite{claret00}):
\begin{equation}
{I(\mu)}/{I(1)} = 1 - \sum_{k=1}^{4}{a_k(1-\mu^{\frac{k}{2}}})
\end{equation}
The $a_k$ coefficient are tabulated by this author for a wide range
of stellar parameters ($T_{\rm eff}$, $\log g$,...) and photometric bands ($U$ to $K$).
For Sirius ($V_T = 0$~km/s, $T_{\rm eff} = 10000$~K, $\log g$ = 4.5, [M/H] = 0)
we find $a_1=0.5318$, $a_2=-0.6921$, $a_3=0.6698$, and $a_4=-0.2431$.
Fig.~\ref{limb_darkening_laws} shows the corresponding intensity profile $I(\mu)/I(1)$.
In this paper, we assume that the limb darkening law does not change over the $K$ band, which is
reasonable for a hot and compact stellar atmosphere, but this computation can easily
be extended to a wavenumber dependent $I(\mu, \sigma)$ intensity profile.
Following Davis et al.~(\cite{davis00}), using a Hankel integral, we can derive the
visibility law $V(B, \theta_{\rm LD}, \sigma)$ from the intensity profile:
\begin{equation}
V =\frac{1}{A} \
{\int_0^1{I(\mu) J_0\left({\pi B\, \sigma \, \theta_{\rm LD} \sqrt{1- \mu^2}} \right)} \mu \ d\mu}
\end{equation}
where $\sigma$ is the wavenumber:
\begin{equation}
\sigma = 1 / \lambda
\end{equation}
and $A$ is a normalization factor:
\begin{equation}
A = {\int_0^1{I(\mu) \mu \, d\mu}} 
\end{equation}

The integral of the binned squared visibilities is computed numerically
over the $K$ band and gives the model $V^2$ for the
projected baseline $B$ and the angular diameter $\theta_{\rm LD}$ through the relation:
\begin{equation}
V^2(\theta_{\rm LD}, B) = \int_K { \left[ V(B, \theta_{\rm LD}, \sigma)\ T(\sigma) \right] ^2 \, d\sigma }
\end{equation}
where $T(\sigma)$ is the normalized instrumental transmission defined so that
\begin{equation}
\int_K T(\sigma)\,d\sigma = 1
\end{equation}
We computed a model of $T(\sigma)$ by taking into account the instrumental
transmission of VINCI and the VLTI. It was first estimated by considering all known
factors (filter, fibers, atmospheric transmission,...) and then calibrated on sky based
on several observations of bright stars with the 8 meter UTs (see Kervella et al.~\cite{kervella03b}
for more details). This gives, for Sirius, an average measurement wavelength of
2.176~$\mu$m.

The $V^2(\theta_{\rm LD}, B)$ model is adjusted numerically to the observed $(B, V^2)$
data using a classical $\chi^2$ minimization process to derive $\theta_{\rm LD}$.
The reduced $\chi^2$ of the global fit is 0.30, characteristic of a satisfactory correspondence between
the model and the measured $V^2$ values. Fig.~\ref{sirius_visib_global} and \ref{sirius_visib_detail}
show the position of the VINCI points on the squared visibility curve of the chosen model.

\subsection{Angular and linear diameters}

The $\chi^2$ minimization gives a limb darkened angular diameter
$\theta_{\rm {LD}} = 6.039~\pm~0.019$~mas for Sirius~A, while a simpler uniform
disk model yields $\theta_{\rm {UD}} = 5.936~\pm~0.016$~mas.
The conversion formula between the LD angular diameter and the
linear diameter $D$ (in solar unit) is:
\begin{equation}\label{distance_eq}
D = \frac{\theta_{\rm {LD}}}{2\,p\ \tan(\theta_{\odot}/2)}
\end{equation}
Assuming a solar angular radius of
$\theta_{\odot}/2 = 959.64~\pm~0.02$'' (Chollet \& Sinceac~\cite{chollet99})
and the {\sc Hipparcos} parallax $p = 379.22 \pm 1.58$~mas,
we obtain $D_{\rm Sirius} = 1.711~\pm~0.013$~D$_{\odot}$.
It should be noted that the parallax uncertainty is largely dominating the final error on
the linear diameter. Therefore, an intrinsically more precise interferometric
measurement would not result in an improvement of the linear diameter precision.
The contribution of the VINCI/VLTI measurement
uncertainty in the error bar is only $\pm 0.004$~D$_{\odot}$, or
37 \% of the total error.

\subsection{Discussion}

Our LD angular diameter value can be compared to the value published by
Hanbury Brown et al.~(\cite{hanbury74a}).
They found, using the Narrabri intensity interferometer,
$\theta_{\rm LD} =$ 5.89 $\pm$ 0.16 mas.
The difference with our value of $\theta_{\rm LD} =$ 6.039 $\pm$ 0.019 mas
is -0.8\,$\sigma$, making the two results statistically compatible within their error bars.

Davis \& Tango~(\cite{davis86}) have obtained a
value of $\theta_{\rm UD} = 5.63 \pm 0.08$\,mas using the amplitude
interferometer, and revised the Hanbury Brown et al.~(\cite{hanbury74a})
value to $\theta_{\rm UD} = 5.60 \pm 0.07$\,mas. Using a limb darkening
coefficient $u = 0.5900$ ($B$ band from Claret et al.\,\cite{claret00}) and
the conversion factor formula  from Hanbury Brown et al.~(\cite{hanbury74b}):
\begin{equation}
\rho = \frac{\theta_{\rm LD}}{\theta_{\rm UD}} = \sqrt{\frac{1-u/3}{1-7u/15}} 
\end{equation}
we find $\rho = 1.053$. This translates into $\theta_{\rm LD} = 5.93 \pm 0.08$\,mas
and $\theta_{\rm LD} = 5.90 \pm 0.07$\,mas for the above UD values, respectively.
They differ respectively by -1.3 and -1.9\,$\sigma$ from our result.
This difference, marginally significant, could be explained by an uncertainty on
the limb darkening factor. It is much stronger at visible wavelengths than
in the infrared, and can be affected by the presence of spectral features
(Tango \& Davis~\cite{tango02}).

Using spectro-photometric observations, Cohen et al.~(\cite{cohen92}) have derived the angular
diameter of Sirius and found 6.04 mas (with Kurucz model atmospheres), in remarkable
agreement with our direct measurement.

Due to the relatively slow rotational velocity of Sirius ($v \sin i~\approx~16$ km.s$^{-1}$,
from Royer et al. \cite{royer02}), we do not expect any detectable flattening of its disk.
In addition, the flux of Sirius~B is totally negligible compared to A,
in particular in the $K$ band.
We therefore do not foresee an asymmetry of the visibility function in
azimuth due to these two contributors.

Hanbury Brown et al.~(\cite{hanbury74a}) have observed a small asymmetry of the Sirius
visibility function, proposing tentatively that this may come from a dust disk around this star.
In the error bars, our measurements do not show this asymmetry, as the D1-C3 and E0-G1 points are
in good agreement with our single disk visibility model. Nevertheless, it should be noted that
we have probed only two azimuth values, namely 80 and 160 degrees
(D1-C3 and E0-G1 baselines, respectively), and that our error bars on the diameter measurements
on the shorter baseline are relatively large.
Therefore, we cannot exclude formally the presence of a disk accounting for approximately
1\% of the stellar flux.

As a remark, Benest \& Duvent~\cite{benest95} have proposed that
a late M5 dwarf could be orbiting Sirius~A, but our measurements do not have the necessary
sensitivity to record the visibility modulation that would be produced by a star
of this type, with $m_K \ge 5$.

\section{Modeling \label{modeling}}
\subsection{Hypothesis}

We computed a number of evolutionary models of Sirius~A using the CESAM
code (Morel~\cite{morel97}) including the pre-main sequence evolution.
The ordinary assumptions of stellar modeling are made, i.e. spherical symmetry,
no rotation, no magnetic field, no mass loss. The relevant nuclear reaction rates
are taken from the NACRE compilation (Angulo et al.~\cite{angulo99}). The equation of
state adopted is EFF (Eggleton et al.~\cite{eff73}), and
the OPAL opacities are from Iglesias \& Rogers (\cite{iglesias96}) with the
Grevesse \& Noels (\cite{gre93}) mixture.
The microscopic diffusion is described using the formalism of 
Burgers~(\cite{burgers69}) with the resistance coefficients of Paquette et al.
(\cite{pa86}). We take into account the radiative diffusivity as
recommended by Morel \& Th\'evenin~(\cite{morel02}), that limits the efficiency
of the microscopic diffusion in outerlayers of stars with intermediate masses.
We have neglected the radiative accelerations, as well as
the changes of abundance ratios between metals within $Z$.
The atmosphere is restored using Hopf's $T(\tau)$ law (Mihalas~\cite{mi1978}). 
The adopted radius of the star is the bolometric radius, 
where $T(\tau_{\star}) = T_{\rm eff}$.
In the convection zones the temperature gradient is
computed according to the $\rm MLT_{CM}$ convection
theory with a mixing length parameter of $\Lambda=1$
(Canuto \& Mazzitelli \cite{cm91}, \cite{cm92}).
Following the prescriptions of Schaller et al.~(\cite{schaller92})
we have computed models that include overshooting of the
convective core (radius $R_{\rm co}$) over the distance
$O_{\rm v}=A\min(H_{\rm p},R_{\rm co})$. We adopted a value A=0.15
in agreement with the results by Ribas, Jordi \& Carme~(\cite{ribas}) for
a 2\,M$_{\odot}$ star.
In addition, we have also computed stellar models without overshooting (A=0.00)
of the convective core and with A=0.20 to estimate the effect of this
parameter on the age of the star.
As a comparison, we have also built a model with enhanced $Z_s/X$,
corresponding to the atmospheric abundances published
by Qiu et al.~(\cite{qiu01}). It is intended to demonstrate the strong effect of
this overestimated value of the internal abundance on the evolution stage of Sirius.
Each model is described by about 760 mass shells. 
We considered that our model was representative of Sirius~A when
we reached the interferometric radius and the observed luminosity and effective
temperature of this star (within their respective error bars).
The main characteristics of the computed models are presented in Table~\ref{tab:glob}.
\subsection{Results and discussion}
%
\begin{figure}[t]
\centering
\includegraphics[width=6.cm,angle=-90]{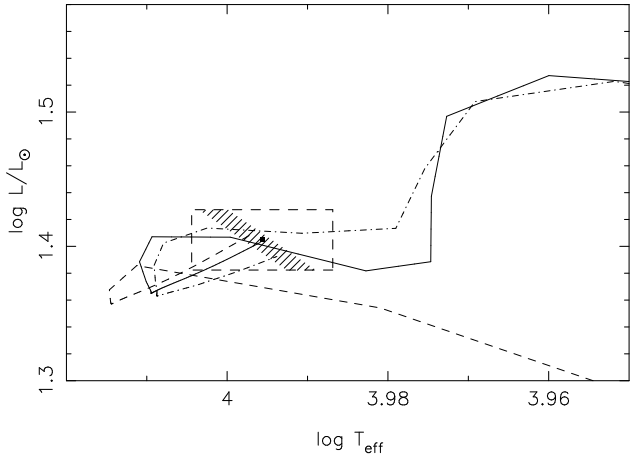}
\caption{
Evolutionary tracks in the HR diagram of models of Sirius A. The dashed rectangle
delimits the uncertainty domain in luminosity and effective temperature, while the shaded area
represents the uncertainty on the interferometric radius. The continuous line corresponds to
model $b$ with overshoot and the dashed-dot line to model $a$ without overshoot,
both with a mass of 2.12 M$_{\odot}$. Model $d$, corresponding to M=2.07
M$_{\odot}$ and an age of 243\,Myr, is represented by a dashed line.
The filled square is the center of the rectangle.
}
\label{HR}
\end{figure}
The HR evolution diagram is
presented on Fig.~\ref{HR}.
We adopt the standard definition of the ZAMS, i. e. where nuclear reactions 
begin to dominate gravitation as the primary source of energy by at least 50\%.
For the mass of Sirius~A this corresponds to an elapsed time of 4.2~Myr.
Note that 100\% of energy comes from nuclear reactions only after 10 Myr.
The evolution of models starts with the quasi-static contraction
of a cloud with a central temperature of 0.5 MK.
According to these definitions, our model $b$ of Sirius, which fits the VLTI/VINCI radius
and satisfies the adopted luminosity and $T_{\rm eff}$, is aged
of $197 \pm 12$~Myr. The error bar is fixed by the uncertainty on the
VINCI/VLTI radius, that is itself largely dominated by the {\sc Hipparcos}
parallax error bar.

Because the age of the WD can reach 210 Myr (Holberg, {\it private communication})
we also computed models with a lower mass 2.07 M${\odot}$ which fit both the
observed radius and the $T_{\rm eff}$. We had to change sligthly $Y_{\rm i}$ and 
the value of $\left(\frac ZX\right)_{\rm i}$ within its error bar, see model $d$,
resulting in an age of 243 Myr for Sirius.
We can produce an derive a greater age of 250 Myr (model $e$) when increasing the overshooting
parameter A to the value A=0.20.
At the end of the pre-main sequence, Sirius~A has generated
a convective core due to the equilibration of the CNO nuclear cycle.
At the expected age of Sirius A for model $d$
the hydrogen mass fraction at center is X$_c$~=~0.590 with a temperature of
22.46~10$^6$~K. The base of the external convective zone is situated at 0.9926 $R_{\star}$
and Y$_s=0.200$. 

Removing the overshooting from our model reduces the age of the binary by 10\%
(model $a$)
while the model $c$ with enhanced abundances leads to a too young age (25 Myr) for
Sirius~A, in significant disagreement with the age of Sirius~B.
At the present status of our observational constraints on Sirius A, it is impossible
to decide if overshoot exists or not in the central convective core.
Note that our model gives a surface abundance of $\rm [M/H]^{Sirius}_{\odot}=-0.32$ dex
which agrees with the abundance of the oxygen found by
Kamp, Hempel \& Holweger~(\cite{kamp02}).
Finally, if Sirius A is able to excite radial modes of oscillation,
we predict the mean large frequency spacing $\overline{\Delta\nu_0}$ to be ranging
between 81 and 82 $\mu$Hz
This is the primary observable of asteroseismology, and corresponds to
the difference between the frequencies of oscillation modes with consecutive
radial order $n$ (see Table~\ref{tab:glob}). This large spacing is sensitive
primarily to the stellar density. It could therefore give a direct, high accuracy
estimation of the mass of Sirius~A when combined with the
observed radius reported in this paper.
\begin{table}
\caption[]
{
Characteristics of our $\alpha$\,CMa\,A models
remaining within the uncertainty box of the
observed properties of this star. The ``Enhanced'' column refers
to the enhanced abundances model.
The subscripts $i$ and $s$ refer respectively to
the initial content at zero age and the surface content at the end of
the computed evolution. The adopted luminosity is
$\log(L/L_{\odot})=1.405\pm 0.022$ with $T_{\rm eff}=9900\pm 200$K.
The observed abundance of oxygen is -0.32$\pm 0.20$ dex (see text).
The precision of the age is $\pm 12$ Myr corresponding to the crossing
time of the shaded parallelogram.
}
\label{tab:glob}
\begin{tabular}{lccccc} \\
\hline
\hline
Model                              &        a      &     b         &      c   &  d     &   e  \\
\hline
M/$M_{\odot}$                      & 2.12          & 2.12          & 2.12     & 2.07   & 2.07 \\
age (Myr)                          & $180$         & $197$         & $25$     & $243$  & $250$\\
overshoot                          & 0.00          & 0.15          &0.15      & 0.15   & 0.20 \\
$T_{\rm eff}$ (K)                  & $9\,870$      & $9\,906$      & $9\,817$ & $9\,936$&$9\,944$\\
$\log (L/L_\odot)$                 &1.398          & 1.403         & 1.385    & 1.408  &1.409\\
$\log\,g$                          & 4.30          & 4.29          & 4.30     & 4.29   &4.29\\
$Y_{\rm i}$                        & 0.267         & 0.267         & 0.330    & 0.273  &0.273\\
$\left(\frac ZX\right)_{\rm i}$    & 0.0165        & 0.0165        & 0.0350   & 0.0145 &0.0145\\
$\left(\frac ZX\right)_{\rm s}$    & 0.0116        & 0.0115        & 0.0289   & 0.0099 &0.0099\\
$\rm[O/H]_{\rm i}$                 & -0.17         & -0.17         &          & -0.23  & -0.23\\
$\rm[O/H]_{\rm s}$                 & -0.33         & -0.33         &          & -0.39  & -0.39\\
$ R/R_\odot$                       & 1.7111        & 1.7107        & 1.7060   & 1.7105 & 1.7103\\
$\overline{\Delta\nu_0}$ ($\mu$Hz) & 82           & 82            & 82       & 81     & 81 \\
 \hline
 \end{tabular}
\end{table}

\section{Conclusion and perspectives}
We have reported in this paper our long-baseline interferometric observations of $\alpha$\,CMa\,A
using the VINCI/VLTI instrument. In conjunction with constraints from previous spectro-photometric
observations, we derived an evolutionary model for Sirius~A.
We propose that the apparently high surface metallic content of Sirius is not characteristic
of the whole average value of $Z$ for the star,  and is caused by the levitation of the
heavy elements on the thin upper convective layer of Sirius A.
By means of our model, we derive an age of 200 to 250$~\pm~12$~Myr, consistent
with the two evolutionary time estimates of Sirius~B.
The oldest age of Sirius~A corresponds to a lower mass of 2.07 M$_{\odot}$
which is the lowest mass value acceptable within its error bar.
 
The accuracy on this age is greatly strengthened by the VLTI/VINCI radius, thus
encouraging further studies to improve our knowledge of the diameter of nearby stars.
Based on our model, we predict that the asteroseismic
large frequency spacing of Sirius~A should be 81 to 82~$\mu$Hz,
if it exhibits radial oscillations.
A more complete modeling of the external layers of this star for asterosismic
frequencies prediction, including radiative accelerations, is currently
in preparation.

We would like to emphasize here the need for very high accuracy parallax values,
beyond the {\sc Hipparcos} precision. We have shown that the uncertainty on our
determination of the linear radius of Sirius~A is largely dominated by the error
on the distance, despite its proximity (2.7~pc).
The next generation of interferometric instruments, such as the AMBER
beam combiner (Petrov et al.~\cite{petrov00}), will provide high precision
angular diameters, together with their wavelength dependence.
In order to constrain the stellar structure models, that output linear values,
it will be necessary to match the precision on the distance to the angular diameter
uncertainty. It is expected that future space based astrometric missions
will be able to give a significant improvement over the {\sc Hipparcos}
parallaxes, and are therefore crucial in this respect.
This work on Sirius~A and the study of $\alpha$\,Cen~A \& B by Kervella
et al.~(\cite{kervella03b}) have demonstrated that for the nearest stars, thanks to
the very small error bar on the linear radius, we have advantage to replace the
classical error bars in the HR diagram, e.g. luminosity and effective
temperature, by the couple luminosity and radius. 

The direct measurement of the limb darkening of Sirius~A is the next step of the
interferometric study of this star. It will allow a refined modeling of its atmosphere
and will complete the calibration of the radius that we presented in this paper.
This observation will be achieved soon using the long baselines of the VLTI.
\begin{acknowledgements}
These interferometric measurements have been obtained using the VINCI instrument
installed at the VLTI. The VLTI is operated by the European Southern Observatory
at Cerro Paranal, Chile. This work has made use of the wavelets data processing technique,
developed by D. S\'egransan (Observatoire de Gen\`eve) and integrated in the VINCI pipeline.
Observations with the VLTI are only made possible through the efforts of the VLTI team,
for which we are grateful. The VINCI public commissioning data reported in this paper have
been retrieved from the ESO/ST-ECF Archive (Garching, Germany).
This research has made use of the SIMBAD database at CDS, Strasbourg (France).
\end{acknowledgements}

{}
\end{document}